\documentclass[11pt]{article}
\usepackage[margin=2.5cm]{geometry}
\usepackage{amsmath,amssymb,extarrows,mathtools,graphicx,subfigure,setspace,bbold}
\usepackage{xcolor}
\usepackage{dsfont}
\usepackage{tikz}
\usetikzlibrary{decorations.pathmorphing}
\usetikzlibrary{decorations.shapes}
\usetikzlibrary{arrows,decorations.pathreplacing}
\usepackage{hyperref}
\usepackage{mathrsfs}
\usepackage{hyperref}
\hypersetup{colorlinks=true, linkcolor=blue, citecolor=magenta, linktoc=page}
\usepackage[square,numbers,sort,compress]{natbib} 
\usepackage{graphicx}

\newcommand{\be}{\begin{equation}}
\newcommand{\ee}{\end{equation}}
\newcommand{\bea}{\begin{eqnarray}}
\newcommand{\eea}{\end{eqnarray}}
\newcommand{\bse}{\begin{subequations}}
	\newcommand{\ese}{\end{subequations}}
\newcommand{\beqa}{\begin{eqnarray}}
\newcommand{\eeqa}{\end{eqnarray}}
\newcommand{\beqar}{\begin{eqnarray*}}
	\newcommand{\eeqar}{\end{eqnarray*}}
\newcommand{\bi}{\begin{itemize}}
	\newcommand{\ei}{\end{itemize}}
\newcommand{\bn}{\begin{enumerate}}
	\newcommand{\en}{\end{enumerate}}

\newcommand{\ba}{\begin{array}}
	\newcommand{\ea}{\end{array}}
\newcommand{\bc}{\begin{center}}
	\newcommand{\ec}{\end{center}}
\newcommand{\nnr}{\nonumber \\}

\newcommand{\eg}{{\em e.g.} }

\def\O{\mathcal{O}}
\DeclareSymbolFont{anttfont}{OML}{antt}{m}{it}
\DeclareMathSymbol{\AFa}{\mathalpha}{anttfont}{`a}
\DeclareMathSymbol{\AFr}{\mathalpha}{anttfont}{`r}
\DeclareMathSymbol{\AFm}{\mathalpha}{anttfont}{`m}
\def\ep{\epsilon}
\def\a{\AFa}
\def\r{\AFr_{+}}
\def\m{\AFm}
\def\l3{\ell}
\def\lb3{\bar{\ell}}

\newcommand{\ev}[1]{\tilde{#1}}
\newcommand{\nh}[1]{\hat{#1}}
\newcommand{\us}[1]{{#1}_{\text{us}}}
\newcommand{\wk}{\us{W}}
\newcommand{\usev}[1]{\us{\ev{#1}}}
\newcommand{\btz}[1]{{#1}_{\text{BTZ}}}

\newcommand{\xdownarrow}[1]{%
	{\left\downarrow\vbox to #1{}\right.\kern-\nulldelimiterspace}}

\begin{document}

\begin{titlepage}

\begin{flushright}\vspace{-3cm}
{
IPM/P-2017/093  \\
\today }\end{flushright}
\vspace{-.5cm}

\begin{center}

\large{\bf{Extremal Vanishing Horizon Kerr-AdS Black Holes at Ultraspinning Limit}}

\bigskip\bigskip

\large{\bf{S.M. Noorbakhsh\footnote{e-mail:
maryam@sciencebeam.com}$^{;\ a}$
and  M.H. Vahidinia\footnote{e-mail: vahidinia@ipm.ir }$^{;\ b}$  }}
\\

\vspace{5mm}
\normalsize
\bigskip\medskip
{$^a$ \it Science Beam Institute, P.O.Box 19848-35367 , Tehran, Iran}
\smallskip

{$^b$ \it School of Physics, Institute for Research in Fundamental	Sciences (IPM),\\ P.O.Box 19395-5531, Tehran, Iran.
 }\\
\date{\today}

\end{center}
\setcounter{footnote}{0}

\begin{abstract}

\noindent

By utilizing the ultraspinning limit we generate a new class of extremal vanishing horizon (EVH) black holes in odd dimensions ($d\geq5$). Starting from the general multi-spinning Kerr-AdS metrics, we show the EVH limit commutes with the ultraspinning limit, in which the resulting solutions possess a non-compact but finite area manifold for all $(t,r\neq r_+)=const.$ slices. We also demonstrate the near horizon geometries of obtained ultraspinning EVH solutions contain an AdS$_3$ throats, where it would be a BTZ black hole in the near EVH cases. The commutativity of the ultraspinning and near horizon limits for EVH solutions is confirmed as well. Furthermore, we discuss only the five-dimensional case near the EVH point can be viewed as a super-entropic black hole. We also show that the thermodynamics of the obtained solutions agree with the BTZ black hole. Moreover we investigate the EVH/CFT proposal, demonstrating the entropy of $2$d dual CFT and Bekenstein-Hawking entropy are equivalent.

\end{abstract}


\end{titlepage}

\tableofcontents
\section{Introduction}  
There has been a great attention in the study of general relativity in higher dimensions, in particular generating new solutions in various contexts with different geometric structures, motivated by their crucial role in the string theory and AdS/CFT. Along these lines, investigation of diverse limits of black holes and black branes have a particular interest. Namely, these limits may lead to new solutions of the same theory and studying their various aspects such as thermodynamics may shed light on some hidden physical features. A more familiar class of asymptotically flat rotating black holes has been constructed by generalization of the known Kerr solution to higher dimensions by Myers and Perry \cite{Myers:1986un}. Its extensions in the presence of the cosmological constant were introduced in four and five dimensions by \cite{Carter:1968ks} and \cite{hep-th/9811056}. Furthermore, the general multi-spinning spinning Kerr-AdS (KAdS) and Kerr-dS solutions in all higher dimension $d$ have been found in \cite{hep-th/0409155,hep-th/0404008}. They also extended their solution by assuming NUT charges in \cite{hep-th/0604125}. It is shown that, for $d\ge 6$, the rotating asymptotically flat black holes may have  arbitrary large angular momentum with the fixed mass by sending the rotation parameter to infinity $(a \to \infty)$ \cite{Myers:1986un}. So they are referred as ultraspinning (US) black holes. 

Not only US limit may be utilized to study instability of rotating black holes at large angular momentum \cite{hep-th/0308056}, but also preforming these limits lead to nontrivial and new interesting geometries. In particular, for asymptotically AdS black holes, a variety of solutions can be generated corresponding with how the rotation parameter $a$ reaches to the AdS radius $l$. For example, by taking $a \to l$ limit, while either the physical mass or horizon radius is kept fixed, the resulting geometry would be a black brane \cite{0806.1954} or hyperboloid membrane  \cite{1206.4351}. In addition, there is another recipe that is applicable to asymptotically (A)dS black holes, in which both $a$ and $l$ go to infinity while  their ratio is finite \cite{1012.5081}.

Recently a new ultraspinning limit was introduced by \cite{1411.4309} such that the generating black hole solutions exhibit a distinguished property which is a noncompact horizon with a finite area. Similar solutions have been found in the context of Einstein-Maxwell-$\Lambda$ theory and $\mathcal{N} = 2$ gauged supergravity coupled to vector multiplets by \cite{1401.3107}. This novel ultraspinning limit can be viewed as a simple generating black hole solution technique in the presence of the cosmological constant. Several authors have applied this method to construct new black hole solutions from  multi-spinning Kerr-AdS black holes (KAdS) \cite{1504.07529}, four and five-dimensional gauged supergravity solutions \cite{1611.02324} and higher dimensional charged AdS black holes \cite{1702.03448}. In all cases, the resulting new exact black hole possess a non-compact horizon with finite entropy, that topologically describes a sphere with some punctures.

Moreover, by examining the behavior of the mentioned new US black holes in the extended thermodynamic phase space, it was shown that they may violate the Reverse Isoperimetric Inequality in some range of parameters \cite{1012.2888}. This inequality implies an upper bound on the horizon area (Bekenstein-Hawking entropy) of a black hole with a certain thermodynamic volume.\footnote{It was shown that (charged) Schwarzschild-AdS black hole has a maximum entropy for a given volume \cite{1012.2888}.} Therefore, since the entropy of these family of US black holes  exceed their corresponding maximum entropy, this limit is also called super-entropic limit \cite{1411.4309,1504.07529}.

Extremality is another interesting limit of black holes and black branes whose surface gravity vanishes. Curiously, in this limit black holes (branes) represent near horizon geometries that not only they belong to a new class of (non-black holes) solutions, but also possesses an enhanced symmetry. For example, extremal $p$-branes in supergravity admit AdS$_{p+1}$ subspace, while the regular extremal black holes generically have an AdS$_2$ sector in the near horizon geometry \cite{hep-th/0606244, 0705.4214}. Interestingly enough, studying them led to AdS/CFT  \cite{hep-th/9711200} and Kerr/CFT (Extremal/CFT), respectively \cite{0809.4266,0811.4393}. There are some uniqueness and existence theorems about the near horizon geometry of extremal black holes \cite{0705.4214,0806.2015,0803.2998,1306.2517,0909.3462}.  Meanwhile, these geometries represent a nice (thermo)dynamic behavior \cite{1310.3727,1407.1992}.  Some features about the near horizon of extremal black holes strongly depend on smoothness of the horizon. However, there is a particular class of them with non-smooth horizon. Among them there is a particular family with vanishing horizon, that their horizon area approach to zero linearly with temperature $A_{H} \propto T \to 0$, near the extremal point. Also the vanishing of the horizon area is a result of vanishing one-cycle on the horizon. Therefore, they are usually  called Extremal Vanishing Horizon (EVH) black holes \cite{hep-th/9905099,0801.4457,0805.0203,0911.2898,1010.4291,1011.1897,1112.4664,1212.3742,1212.4553,1301.3387,1301.4174,1308.1478,1407.7484,1510.01209,1703.01771}.

Near horizon limit of EVHs also provides a new class of solutions and they enjoy an enhanced symmetry. It has been proved that the EVH black holes in Einstien-Maxwell-Dilaton-$\Lambda$ gravity admit a three-dimensional maximally symmetric subspace in the near horizon. In the case of Einstein-$\Lambda$ with negative cosmological constant, this subspace would be an AdS$_3$. For a near EVH (NEVH) black hole, where the entropy and the temperature are small but non zero, the AdS$_3$ will be replaced by a BTZ black hole \cite{1107.5705,1504.03607,1512.06186}. Recently, it was shown in \cite{1703.01771} that in the case of EVH Kerr-dS black holes, this $3$d part can interpolate between dS$_3$, flat and AdS$_3$ spaces depending on rotation parameters, while the near horizon of an EVH cosmological soliton always present a dS$_3$. Appearing the AdS$_3$ sector in the near horizon of EVH solutions suggests that low energy physics on the EVH geometries may have a dual holographic description in terms of a CFT$_2$. This duality is usually called EVH/CFT correspondence \cite{1107.5705}.

\begin{figure}[t]
	\centering
	\includegraphics[scale=1]{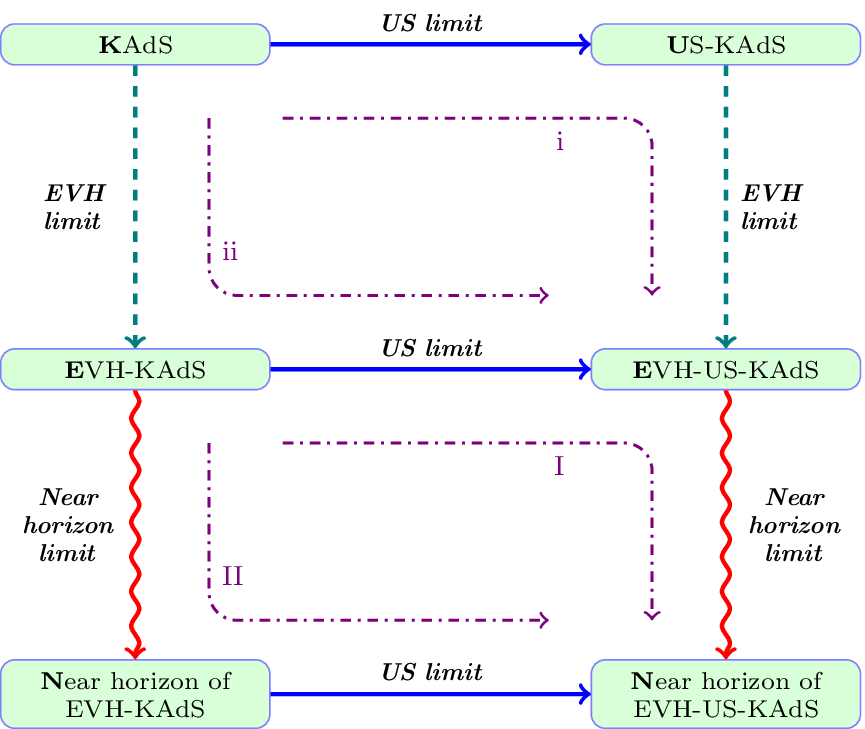}
	\caption{Different order of limits for KAdS black hole}	\label{Fig1}
\end{figure}

The interesting features of EVH and ultraspinning black holes, motivated us to mix EVH and US limits on a given metric by taking different orders. The primary purpose of this work is investigating if the ultraspinning black holes have a well-defined EVH description, and exploring EVH configurations from  ultraspinning black holes. Since EVH black holes suffer a singularity on the horizon, we also were curious to examine whether the US limit can lead to getting rid of this problem. For these purposes, we will consider the general  multi-spinning Kerr-AdS (KAdS) black holes in all dimensions. Then by examining both of these limits in different orders, as illustrated in Fig. \ref{Fig1}, we will show that by following both paths  one can generate a single new geometry, which is the EVH ultraspinning KAdS (EVH-US-KAdS) black hole in odd dimensions.

Moreover, as it is shown in the lower part of Fig. \ref{Fig1}, we investigate not only EVH and US limits commute with each other, but also the near horizon of EVH-US-KAdS black hole can be obtained by performing an appropriate ultraspinning limit on the near horizon of the EVH-KAdS solution. We also explore the existence of a dual CFT by following EVH/CFT to find a low energy description for EVH at large angular momentum.

This paper is organized as follows. In Sec. \ref{SecKAdS} we review the general KAdS black hole, followed by its US and (N)EVH limits as well as the near horizon of (N)EVH cases. In Sec. \ref{SecUSEVH} we show that US-KAdS has an EVH configuration, that can be reproduced by applying ultraspinning limit on EVH solution, and vice versa. In Sec. \ref{SecNH} we present the near horizon geometry of the obtained solutions by different strategies. Furthermore, in Sec. \ref{SecThermo} we discuss the thermodynamics of BTZ black hole, which is located in the near horizon geometry of our NEVH-US solutions in Sec. \ref{SecSuper}. We also explore a super-entropic EVH solution in Sec. \ref{SecSuper}. The Sec. \ref{SecConclusion} is devoted to  conclusions with some remarks. In appendix \ref{App1} we discuss on the topology of the obtained EVH-US solutions.

\section{Various limits of multi-spinning Kerr-AdS}\label{SecKAdS}
We start with a brief review of ultraspinning and (near) EVH limits on multi-spinning spinning Kerr-AdS (KAdS) black holes separately, followed by their construction. Consider the general multi-spinning spinning Kerr-AdS black holes in $d$ spacetime dimensions \cite{hep-th/0409155,hep-th/0404008}. Such metrics are a generalization of asymptotically flat higher dimensional Myers-Perry black hole solutions \cite{Myers:1986un}, and represent the general solutions of vacuum with negative cosmological constant
\begin{equation}\label{EinsteinLambda}
R_{ab}=-\frac{d-1}{l^2}g_{ab}
\end{equation}
The metric in the generalized Boyer-Lindquist coordinates is given by \footnote{The more general Kerr-NUT solution is presented in \cite{hep-th/0604125}}
\begin{equation}\label{Metric1}
ds^2= d\gamma^2 + \frac{2m}{U} \omega^2 + \frac{U}{X}dr^2+d \Omega^2,
\end{equation}
where 
\begin{gather}\label{Metric2}
d\gamma^2=-\frac{W \rho^2}{l^2}dt^2 + \sum_{i=1}^{N} \frac{r^2+a_i^2}{\Xi_i}\mu^2_i d \phi^2_i, \quad \omega=W dt - \sum_{i=1}^{N}\frac{a_i \mu^2_i d \phi_i}{\Xi_i},\\ \nonumber
d\Omega^2= \sum_{i=1}^{N+\delta}\frac{r^2+a_i^2}{\Xi_i}d \mu^2_i - \frac{1}{W \rho^2}\bigg(\sum_{i=1}^{N+\delta}\frac{r^2+a_i^2}{\Xi_i} \mu_i d\mu_i\bigg)^2,\\ 
X=\frac{r^{\delta-2} \rho^2}{l^2} \prod_{i=1}^{N}(r^2+a^2_i)-2m, \quad\quad \rho^2=r^2+l^2, \quad\quad \Xi_i=1-\frac{a^2_i}{l^2}. \nonumber \\
W= \sum_{i=1}^{N+\delta}\frac{\mu^2_i}{\Xi_i}, \qquad\qquad U= r^\delta \sum_{i=1}^{N+\epsilon}\frac{\mu^2_i}{r^2+a^2_i}\prod_{j}^{N}(r^2+a^2_j),\nonumber
\end{gather}
and it is assumed $\delta=1, 0$, for even and odd dimensions, respectively. In general, in this spacetime we are dealing with $N=[(d - 1)/2]$ azimuthal coordinates $\phi_i$ corresponding to $N$ independent rotation parameters $a_i$. Particularly, $a_{N+1}=0$, in even dimensions. Moreover, there are $[d/2]$ polar coordinates or ``direction cosines" $\mu_i$, that are related to each other by the following constraint
\begin{equation}
\sum_{i=1}^{N} \mu^2_i=1.
\end{equation}
The metric Eq. \eqref{Metric1} describes a black hole if $X(r_{+})=0$, allows a real root that determines the location of the black hole horizon. Let $r_+$ denotes the largest root as an outer horizon, which is generated by the Killing vector $K=\partial_t+\sum_{i}^{N} \Omega_i^S \partial_{\phi_i}$, where $\Omega_i^S$'s are $N$ independent angular velocities on the horizon in the asymptotic static frame (ASF) as
\begin{eqnarray}\label{OmegaS}
\Omega_i^{\text{S}}=\frac{a_i(1+r_+^2/l^2)}{r_+^2+a_i^2},
\end{eqnarray}
Note, the metric Eq.  (\ref{Metric1}) is already written in the ASF. The corresponding angular velocities in the asymptotic rotating frame (ARF) read
\be\label{OmegaR}
\Omega_i^{\text{R}}=\Omega_i^{\text{S}}-\frac{a_i}{l^2}.
\ee
The Hawking temperature and entropy are given by \cite{hep-th/0409155}
\begin{gather}\label{Thermo}\nonumber
S=\frac{A_{H}}{4 G}=\frac{\mathcal{A}_{d-2}}{4Gr_+^{1-\delta}}\prod_{i=1}^{N}\frac{a_i^2+r_+^2}{\Xi_i}, \\
T=\frac{\kappa_{H}}{2 \pi}=\frac{1}{2\pi}\bigg[r_+\bigg(\frac{r_+^2}{l^2}+1\bigg)\sum_{i=1}^{N}\frac{1}{a_i^2+r_+^2}-\frac{1}{r _+}\bigg(\frac{1}{2}-\frac{r_+^2}{2 l^2}\bigg)^{\delta}\,\bigg],
\end{gather}
where $\mathcal{A}_{d}$ is the volume of a unit $d$-sphere
\begin{gather}\label{A-unit}\nonumber
\mathcal{A}_{d}=\frac{2\pi^{(d+1)/2}}{\Gamma[(d+1)/2]}.
\end{gather}
The conserved charges of this solution, including mass and $J_i$ angular momenta read \cite{hep-th/0408217}
\begin{eqnarray}\label{Charges}
M=\frac{m \mathcal{A}_{d-2} }{4 \pi G (\prod_j \Xi_j)}\big(\sum_{i=1}^{N}\frac{1}{\Xi_i} - \frac{1-\delta}{2}\big), \quad \quad \quad J_i=\frac{a_i m \,\mathcal{A}_{d-2}}{4 \pi G \Xi_i (\prod_j \Xi_j)}.
\end{eqnarray}
One can check these quantities satisfy the first law of thermodynamics.

In what follows, we shall present the ultraspinning and EVH limits of metric Eq.  (\ref{Metric1}) separately.
\subsection{Ultraspinning limit}\label{US-limit}
We proceed by briefly recapitulating the novel ultraspinning limit of Kerr-AdS metric (US-KAdS) Eq. \eqref{Metric1} that is introduced in \cite{1411.4309,1504.07529}. As it was mentioned in the introduction, this limit can be utilized as a simple generating solution method in the presence of negative cosmological constant. In which one of the rotation parameter that we choose $a_{k}$, reaches to its maximum value that is equal to the AdS radius $l$.
 
This novel technique is performed in three steps \footnote{This new ultraspinning limit is dubbed super-entropic as well. For more details see \cite{1411.4309,1504.07529}}:
i) transforming a given Kerr-AdS black hole to an asymptotically rotating frame (ARF). Then to avoid any possible divergences in the metric, it is crucial to define a new rescaled azimuthal coordinate $\varphi_k=\phi_k/{\Xi_k}$,  ii) boosting the corresponding asymptotic rotation to the speed of light, easily by taking $a_k \rightarrow l$ limit, iii) compactifying the new azimuthal coordinate $\varphi_k$.
 
The ultraspinning version of general Kerr-AdS solutions Eq.  (\ref{Metric1}) has been  investigated in \cite{1504.07529}. It should be noted that, this ultraspinning limit is applicable only in one azimuthal direction $\phi_{k}$. Indeed, as it was argued in \cite{1504.07529}, it is impossible to apply more than one ultraspinning limit simultaneously. 

Before following the above steps towards obtaining the ultraspinning limit of Eq. \eqref{Metric1}, we choose the ultraspinning limit in an arbitrary $\phi_{k}$ direction. Then, it would be necessary to transform Eq. \eqref{Metric1} to ARF by setting
\begin{equation}
\phi_k=\phi_k^R+\frac{a_k}{l^2} t.
\end{equation}
Now, by introducing a new coordinate 
\begin{equation}
\varphi_k=\frac{\phi_k^R}{\Xi_k},
\end{equation}
and finally upon taking the limit $a_k \rightarrow l$, the ultraspinning version of Eq.  (\ref{Metric1}) can be obtained  \cite{1504.07529}
\begin{equation}\label{USMetric1}
ds^2= d\us{\gamma}^2 + \frac{2m}{\us{U}} \us{\omega}^2 + \frac{\us{U}} {\us{X}}dr^2+d \us{\Omega}^2,
\end{equation}
where 
\begin{eqnarray}\label{USMetric2}\nonumber
d\us{\gamma}^2&=&-\bigg((\wk+\mu_k^2)\rho^2+\mu_k^2 l^2\bigg)\frac{dt^2}{l^2} +\frac{2\rho^2 \mu_k^2 dt d\varphi_{k}}{l}  + \sum_{i \neq k}^{N} \frac{r^2+a_i^2}{\Xi_i}\mu^2_i d \phi^2_i,\\ \nonumber
d\us{\Omega}^2&=&\sum_{i \neq k}^{N+\delta}\frac{r^2+a_i^2}{\Xi_i}d \mu^2_i -2 \frac{d\mu_k}{\mu_k}\bigg(\sum_{i \neq k}^{N+\delta}\frac{r^2+a_i^2}{\Xi_i} \mu_i d \mu_i\bigg)+\frac{d \mu^2_k}{\mu_k^2}\bigg(\rho^2\wk+l^2\mu_k^2\bigg),\\ 
\us{\omega}&=&(\wk+\mu_k^2)dt -l \mu_k^2 d\varphi_k- \sum_{i \neq k}^{N+\delta}\frac{a_i \mu^2_i d \phi_i}{\Xi_i}, \qquad \qquad \wk= \sum_{i \neq k }^{N+\delta}\frac{\mu^2_i}{\Xi_i}, \nonumber \\
\us{X}&=&\frac{r^{\delta-2} \rho^4}{l^2} \prod_{i \neq k}^{N}(r^2+a^2_i)-2m, \qquad \us{U}= r^\delta\bigg(\mu_k^2 + \sum_{i \neq k}^{N+\delta}\frac{\mu^2_i \rho^2}{r^2+a^2_i}\bigg)\prod_{s \neq k}^{N}(r^2+a^2_k).
\end{eqnarray}
Also the new coordinate $\varphi_k$ requires to compactify as
\begin{equation}\label{mu}
\varphi_k \sim \varphi_k + \mu.
\end{equation}

This geometry describes a new class of exact black hole solutions to the Einstein-$\Lambda$ theory in all higher dimensions. Interestingly, it was shown that the horizon of the above solution topologically is a sphere with some punctures, coming from $\mu_k=0$ poles. That gives rise to a non-compact horizon black hole. Indeed, the poles $\mu_k=0$ are removed from the spacetime, and can be viewed as a kind of boundary \cite{1504.07529}. Nevertheless, the area of the horizon remains finite and entropy is given by
\begin{gather}\label{Area-US}
S=\frac{\bar{\mathcal{A}}_{d-2}}{4 G r_+^{1-\delta}}\rho_+^2\prod_{i \neq k}^{N}\frac{a_i^2+r_+^2}{\Xi_i},
\end{gather}
where 
\begin{eqnarray}\label{Amu}
\bar{\mathcal{A}}_d=\frac{\mu\,\pi^{(d-1)/2}}{\Gamma[(d+1)/2]}=\frac{\mu}{d-1}\mathcal{A}_{d-2}.
\end{eqnarray}
and $\mu$ denotes the dimensionless parameter which is introduced in Eq. \eqref{mu} for compaticyifing the new coordinate $\varphi_k$. 
In \cite{1504.07529} it is argued that $\mu$ can be interpreted as a thermodynamic parameter, relating to a chemical potential. However, we take it as a constant through this paper.

\textbf{Thermodynamic quantities:}
For the ultraspinning solution  Eq.  (\ref{USMetric1}), the thermodynamic quantities, including mass, angular momenta, Hawking temperature and angular velocities on the horizon are computed \cite{1504.07529}
\be\label{Thermo-US}
\begin{gathered}
T=\frac{1}{2\pi}\bigg[\frac{r_+}{l^2}\bigg(1+\sum_{i \neq k}^{N}\frac{l^2+r_+^2}{a_i^2+r_+^2}\bigg)-\frac{1}{r_+}\bigg(\frac{1}{2}-\frac{r_+^2}{2l^2}\bigg)^{\delta}\,\bigg],\\  
\Omega_k= \frac{l}{l^2+r_+^2}, \qquad \qquad \Omega_{i \neq k}= \frac{a_i\,(l^2+r_+^2)}{l^2(r_+^2+a_i^2)},\\   
M=\frac{m \, \tilde{\mathcal{A}}_{d-2} }{4 \pi G \prod_{j \neq k} \Xi_j}\big(\sum_{i \neq k}^{N}\frac{1}{\Xi_i}+ \frac{1+\delta}{2}\big), \\
J_k=\frac{l \, m \, \tilde{\mathcal{A}}_{d-2}}{4 \pi G \prod_{k \neq j} \Xi_j},  \qquad  \qquad J_{i \neq k}=\frac{a_i \,m\, \tilde{\mathcal{A}}_{d-2}}{4 \pi G \Xi_i \prod_{j \neq k} \Xi_j}.
\end{gathered}
\ee

Notice that, these conserved charges cannot be obtained simply by taking the limit $a \to l$ on thermodynamic quantities of origin general Kerr-AdS black hole Eq. \eqref{Metric1}. Although, there is no trivial relation between each of the thermodynamic quantities of US solution Eq. \eqref{USMetric1} and their correspondence to the original black hole Eq. \eqref{Metric1}, it is straightforward to check that the first law of thermodynamics is preserved under the US limit.

\subsection{EVH and near-EVH limits}\label{subSecEVH22}
The EVH black hole solutions describe a class of extremal black holes $(T=0)$  with vanishing horizon area $A_H$, while keeping the ratio $A_H/T$ finite \cite{1107.5705}. It is usually assumed the vanishing horizon area $A_H$ is due to a vanishing one-cycle on the horizon. Namely, the metric of the horizon has only one zero eigenvalue. The definition is as follows
\be \label{EVH}
A_H, T \sim \epsilon \rightarrow 0, \qquad  \frac{A_H}{T}=\text{fixed}.
\ee
This particular family of EVH black holes enjoys a three-dimensional maximally symmetric subspace in the near horizon geometry. It was shown that in the generic Einstein-Maxwell-dilaton-$\Lambda$ theories with negative cosmological constant $(\Lambda<0)$, the three-dimensional near horizon region is an AdS$_3$ geometry  \cite{1512.06186,1504.03607}. 

As we shall show, only in odd dimensions one can construct an EVH solution from the metric Eq.  (\ref{Metric1}), obeying the one-cycle EVH definition Eq.  (\ref{EVH}). However, in the case of even dimensions, one may find a new class of EVH solutions by imposing $A_{H}/T^2$ remains finite. Though it could be an interesting research, but it is beyond scope of the current paper.

Therefore, in the rest of this paper we only consider odd dimensions. The EVH limit of Kerr-AdS black hole in odd dimensions has been studied in details in \cite{1703.01771} \footnote{Similar study has been done for asymptotically flat five-dimensional Myers-Perry black hole  \cite{1308.1478}}. Here, we just focus on the main results for our purpose.

\textbf{EVH limit:} Following \cite{1703.01771}, to find an EVH limit of metric Eq.  (\ref{Metric1}), one can proceed two steps: first, we turn off one of the rotation parameters along an arbitrary direction, namely $\phi_p$, by setting $a_p=0$. Then we set $r_{+}=0$. In other words
\be \label{EVHlimit}
r_{+} \sim \epsilon,  \qquad  a_{p} \sim \epsilon^2, \qquad m=\frac{1}{2}\prod_{i \neq p}^{N}a_{i}^2, \qquad \ep \to 0,
\ee 
where the constraints between $m$ and $a_i$'s come from the consistency among $X(r_{+})=0$, as well as scaling of $r_{+}$ and $a_{p}$. It is straightforward to check that $A_{H}/T$ remains finite in the above configuration.

\textbf{Near-EVH limit:}
The EVH black hole is described by a codimension-two surface in $n+1$ dimensional parameter space of the KAdS solution. Along this surface both temperature and horizon area vanish. It is also interesting to explore physics of black hole near this EVH surface, where black hole has a small non-zero temperature and entropy. So, we define the \emph{near}-EVH limit precisely as
\be \label{NearEVHlimit}
r_{+}=\r \;\epsilon,  \qquad  a_{p}=\a \; \epsilon^2, \qquad m=\frac{1+\m \; \ep^2}{2}\prod_{i \neq p}^{N}a_{i}^2, \quad \ep \to 0,
\ee 
 where $\r$ and $\a$ are two dimension-full parameters. Indeed, $\m\ep^2$ denotes a small deviation from the EVH surface, and it is given by 
\begin{gather}\label{Delta_m}
\m=  \frac{\a^2}{\r^2}+\frac{\r^2}{\lb3^{2}} , \quad \lb3^{-2}\equiv \frac{1}{l^2}+\sum_{i \neq p}^{N}a_{i}^{-2}.
\end{gather}

The above \emph{near}-EVH limit Eq. \eqref{EVHlimit} leads to the following quantities for temperature and area of the horizon 
\begin{eqnarray}\label{TAevh}
T=\frac{1}{2\pi}(\frac{1}{\lb3^2}-\frac{\a^2}{\r^4})\,\r \ep, \qquad A_{H}=\mathcal{A}_{d-2}\left(\prod_{i \neq p}^{N}\frac{a_{i}^2}{\Xi_{i}}\right)\;  \r \ep,
\end{eqnarray}
The ratio $A_{H}/T$ remains finite at the EVH limit ($\ep \to 0$). We can also apply (near) EVH limit to angular velocity and angular momentum along $\phi_{p}$ direction
\be 
\Omega_{p}^{\text{S}}=\frac{\a}{\r^2}+\mathcal{O}(\ep^2), \qquad \qquad J_{p}=\frac{\mathcal{A}_{d-2}}{8 \pi G}\prod_{i \neq p}^{N}\frac{a_{i}^2}{\Xi_{i}} \;\a \; \ep^2+\mathcal{O}(\ep^4).
\ee

It is worthwhile to mention, consistency for the existence of a black hole solution ($X(r_{+})=0$), gives rise to a further constraint that only one of $a_{i}$'s is allowed to vanish \eg $a_{p}$. Thus, there is no static EVH configurations for general multi-spinning Kerr-AdS black holes Eq.  (\ref{Metric1}). While for the most previous EVH examples in the literature, such as five-dimensional $U(1)^3$ gauged supergravity black holes, one can find a static EVH configuration \cite{1301.3387}.
	
\subsubsection{EVH Kerr-AdS black holes}
To present the explicit description of the EVH solution, we apply the EVH conditions $a_p=r_+=0$ Eq. \eqref{EVHlimit} onto the metric Eq.  (\ref{Metric1}).
Straightforwardly one can obtain the EVH geometry 
\begin{equation}\label{EVHMetric1}
d\ev{s}^2=d\ev{\gamma}^2+\frac{2m}{\ev{U}}\ev{\omega}^2+\frac{\ev{U}}{\ev{X}}dr^2+d\ev{\Omega}^2
\end{equation}
where 
\begin{eqnarray}\label{MetricEVH}\nonumber
d\ev{\gamma}^2&=&-\frac{\ev{W} {\rho}^2}{l^2}dt^2 + \sum_{i \neq p}^{N} \big( \frac{r^2+a_i^2}{\Xi_i}\mu^2_i \big) d \phi^2_i+r^2 \mu^2_p d\phi^2_p, \\ \nonumber
d\ev{\Omega}^2&=& r^2 d\mu_p^2 + \sum_{i \neq p}^{N}\frac{r^2+a_i^2}{\Xi_i}d \mu^2_i - \frac{1}{\ev{W} \rho^2}\big(  r^2 \mu_p\, d\mu_p + \sum_{i \neq p}^{N}\frac{r^2+a_i^2}{\Xi_i} \mu_i d\mu_i\big)^2 ,\\ 
\ev{U}&=&r^2 \big(\sum_{i \neq p}^{N }\frac{\mu^2_i}{r^2+a^2_i} + \frac {\mu_p^2}{r^2}\big)\prod_{j \neq p}^{N}(r^2+a^2_j), \qquad \ev{W}= \sum_{i \neq p }^{N}\frac{\mu^2_i}{\Xi_i}+\mu_{p}^2,  \nonumber \\
\ev{\omega}&=&\ev{W} dt - \sum_{i \neq p}^{N}\frac{a_i \mu^2_i d \phi_i}{\Xi_i}, \qquad \qquad \qquad  \ev{X}=\frac{\rho^2}{l^2} \prod_{i \neq p }^{N}(r^2+a^2_i)-\prod_{i \neq p}^{N}a_{i}^2.
\end{eqnarray}
where ``tilde" components refer to those expressions in Eq.  (\ref{Metric2}) that are calculated exactly at the EVH point.


\subsubsection{Near horizon of EVH Kerr-AdS}
One can construct the near horizon geometry of the EVH-KAdS black hole Eq. \eqref{MetricEVH}, via performing the following dimensionless coordinate transformation \cite{1703.01771}
\begin{gather}\label{NearHorizonLimit}
r=r_{+}+\lambda \; \nh{r} , \qquad t= \frac{\nh{t}}{\lambda }, \qquad \phi_{p} = \frac{\nh{\phi}_p}{\lambda },\qquad \phi_{i \neq p}= \nh{\phi}_{i}+\Omega_{\text{S}}^{i} \; \nh{t},\qquad \hat{\mu_i}=\mu_i,
\end{gather}
followed by the limit $\lambda \rightarrow 0$, while we set the horizon radius $r_{+}=0$. Ultimately, we find the near horizon metric as
\be\label{NearHorizonEVH}
ds_{\text{N.H.}}^2=\mu_{p}^2\left(-\frac{\nh{r}^2}{\lb3^2}d\nh{t}^2+\frac{\lb3^2}{\nh{r}^2}d\nh{r}^2+\nh{r}^2 d\nh{\phi}_{p}^2\right)+\sum_{i,j \neq p}{(h_{ij}d\mu_{i}d\mu_{j}+k_{ij}d\nh{\phi}_{i}d\nh{\phi}_{j})}
\ee
where 
\bea
h_{ij}&=&\frac{a_i^2\,\mu_i^2}{\Xi_i}\, \delta_{ij}+\frac{\mu_i^2\, \mu_j^2}{\mu_p^2} \frac{a_i\, a_j }{\Xi_i\,\Xi_j} \,,\nnr\nnr
k_{ij}&=&\frac{a_i^2}{\Xi_i}\; \delta_{ij}-\,\frac{1}{l^2}\; \frac{\mu_i\, \mu_j}{\ev{W}}\, \frac{a_i^2\, a_j^2}{\Xi_i \Xi_j} \,.
\eea
We note, this metric satisfies $d$-dimensional Einstein equation in the presence of negative cosmological constant Eq.  (\ref{EinsteinLambda}). Clearly, the three-dimensional part of $(\nh{t},\nh{r},\nh{\phi}_{p})$ represents an AdS$_{3}$ spacetime with radius $\lb3$ is given by Eq. \eqref{Delta_m}.  It should be noted, due to the near horizon limit Eq. \eqref{NearHorizonLimit}, the periodicity of $\nh{\phi}_{p}$ is reduced to $2\pi\lambda $. Thus, this three-dimensional space can be viewed as a \emph{pinching} AdS$_3$ \cite{1011.1897,1107.5705}. In addition, we note that this geometry is regular everywhere except at $\mu_{p}=0$, where Kretschmann scalar diverges. This usually happens in EVH black holes \cite{1107.5705,hep-th/9905099} (for a counter example see \cite{1703.01771}).

\subsubsection{Near horizon of near-EVH Kerr-AdS}
It was shown that the near horizon geometry of a near-EVH black hole admits a more general metric. To obtain this geometry one should take the near-EVH limit Eq. \eqref{NearEVHlimit} and the near horizon limit Eq. \eqref{NearHorizonLimit} simultaneously. After taking both limits and assuming $\ep \sim \lambda $, one finds
\be\label{NearHorizonNearEVH}
ds_{\text{N.H.}}^2=\mu_{p}^2 \; ds_{\text{BTZ}}^2+\sum_{i,j \neq p}{(h_{ij}d\mu_{i}d\mu_{j}+k_{ij}d\nh{\phi}_{i}d\nh{\phi}_{j})}
\ee
where $ds_{\text{BTZ}}^2$ denotes a pinching BTZ black hole as
\begin{gather}\label{BTZ}
ds_{\text{BTZ}}^2=-f(\nh{r})d\nh{t}^2+\frac{d\nh{r}^2}{f(\nh{r})}+\nh{r}^2\left(d\nh{\phi}_{p}-\frac{\a}{\nh{r}^2}d\nh{t}\right)^2, \nonumber \\
f(\nh{r})=\frac{(\nh{r}^2-\r^2)(\nh{r}^2-\;\frac{\a^2\;\lb3^2}{ \r^2})}{\l3^2 \; \nh{r}^2 }.
\end{gather}
This geometry reduces to near horizon of EVH-KAdS
Eq.  Eq.  \eqref{NearHorizonEVH} in the case of $\a=0=\r$.
It was also argued that the thermodynamics of this BTZ black hole are corresponding to thermodynamics of the near EVH-KAdS black hole
\cite{1107.5705,1301.3387,1703.01771}.

\section{Ultraspinning and EVH limits}\label{SecUSEVH}
In this section we shall show the ultraspinning limit can be combined with the EVH limit, allowing us to generate a new class of EVH solutions at ultraspinning limit. There are two ways towards constructing this new type of solutions from the general Kerr-AdS black holes Eq.  (\ref{Metric1}). (i) beginning with their EVH geometries and then performing the ultraspinning limit, (ii) beginning with the ultraspinning version and then finding their EVH solutions afterwards. As Fig. \ref{Fig1} shows these two different procedures yield the identical EVH ultraspinning KAdS (EVH US-KAdS). 

\begin{figure}[t]
\begin{center}
\includegraphics[scale=1]{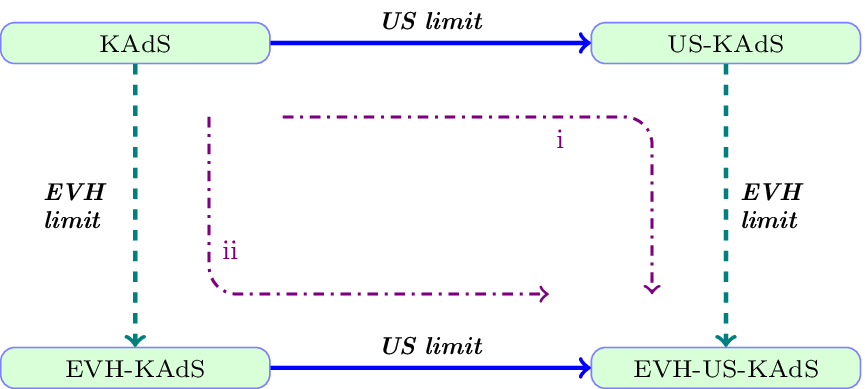}
\caption{This diagram shows two different order limits for KAdS black hole. Horizontal arrows (blue) represent the ultraspinning (US) limit. The vertical ones (green) show the EVH limit. We show that in both paths the resulting geometries (US-EVH-KAdS) are exactly the same.}\label{Fig2}
\end{center}
\end{figure}
\subsection{ Ultraspinning limit of EVH black hole }
Let us first study the lower path in Fig. \ref{Fig1}, which begins with the EVH solution that is already derived in Eq. \eqref{EVHMetric1}, and then construct its ultraspinning version by following the method of \cite{1504.07529} that has been summarized in Sec. \ref{US-limit}.

To start the procedure towards an ultraspinning EVH solution, one has to choose an arbitrary $\phi_k$ coordinate as an ultraspinning direction. Before applying the ultraspinning limit to the whole geometry, let us use a trick, which is separating the $\phi_k$ from the other $\phi_{i \neq k}$ coordinates. Therefore, we easily find
\begin{equation}
\tilde{W} =\frac{\mu_k^2}{\Xi_k}+\mu_p^2 + \sum_{i \neq p \neq k}^{N}\frac{ \mu_i^2}{\Xi_i}.
\end{equation}
now by taking the limit $a_k\rightarrow l$, the following result can be obtained
\begin{equation}\label{US-1}
\tilde{W}\, \Xi_k = \mu_k^2.
\end{equation}
Using the above result, the $d\ev{\Omega}^2$ part of the metric \eqref{EVHMetric1} takes the following form
\begin{eqnarray}\label{OmegaUS2}
d\ev{\Omega}^2&=& r^2 d\mu_p^2 + \sum_{i \neq p \neq k}^{N}\frac{r^2+a_i^2}{\Xi_i}d \mu^2_i - \frac{\Xi_k}{\ev{W} \,\Xi_k\,\rho^2}\big(  r^2 \mu_p\, d\mu_p + \sum_{i \neq p \neq k}^{N}\frac{r^2+a_i^2}{\Xi_i} \mu_i d\mu_i\big)^2 \\ \nonumber
&-&  \frac{2 (r^2+a_k^2)}{\ev{W} \,\Xi_k\,\rho^2}\mu_k d\mu_k\big(r^2 \mu_p\, d\mu_p + \sum_{i \neq p \neq k}^{N}\frac{r^2+a_i^2}{\Xi_i} \mu_i d\mu_i\big) + \frac{r^2+a_k^2}{\Xi_k}d\mu_k^2\bigg(1-\frac{r^2+a_k^2}{\ev{W}\,\Xi_k\,\rho^2}\mu_k^2\bigg).
\end{eqnarray} 
Now, the limit $a_k \rightarrow l$ can be easily taken for the first four terms, while the last term needs some calculation as follow
\begin{eqnarray}\label{OmegaUS3}\nonumber
& &\frac{r^2+a_k^2}{\tilde{W}\Xi_k} d\mu_k^2 \big(\tilde{W}-\frac{\rho^2-\Xi_k\,l^2}{\Xi_k \rho^2} \mu_k^2\big) \\ 
&=& \frac{r^2+a_k^2}{\tilde{W}\Xi_k} d\mu_k^2  \big(\mu_p^2+ \sum_{i \neq p \neq k}^{N} \frac{\mu_i^2}{\Xi_i} + \frac{l^2}{ \rho^2} \mu_k^2\big).
\end{eqnarray}
Now, Putting everything together and using the result Eq.  (\ref{US-1}), by sending $a_k \rightarrow l$, we finally obtain $d\ev{\Omega}^2 \rightarrow d\usev{\Omega}^2$ as follow
\begin{eqnarray}\label{OmegaUSEVH}\nonumber
d\usev{\Omega}^2&=& r^2 d\mu_p^2+ \sum_{i \neq p \neq k}^{N}\frac{r^2+a_i^2}{\Xi_i}d \mu^2_i -2 \frac{d \mu_k}{ \mu_k} \big(r^2\mu_p d\mu_p+\sum_{i \neq j \neq k}^{N}\frac{r^2+a_i^2}{\Xi_i} \mu_i d \mu_i\big)^2\\ 
&+&\frac{d\mu_k^2}{\mu_k^2} \big( \rho^2\usev{W} + l^2\mu_k^2\big),
\end{eqnarray}
where $\usev{W}$ reads
\begin{equation}\label{Ws}
\usev{W}= \mu_p^2 + \sum_{i \neq p \neq k}^{N} \frac{\mu_i^2}{\Xi_i} .
\end{equation}
Now, we are ready to proceed the ultraspinning limit. We start by following coordinate transformation to switch the ARF
\begin{equation}\label{ChangePhi}
\phi_k=\phi_k^R+\frac{a_k}{l^2}t.
\end{equation}
Then we have
\begin{eqnarray}
\ev{\omega}&=&(\usev{W}+\mu_k^2)dt - \frac{a_k \mu_k^2 d\phi_k}{\Xi_k} - \sum_{i \neq p \neq k}\frac{a_i \mu_i^2 d\phi_i}{\Xi_i}, \\ \nonumber
d\ev{\gamma}^2&=&\big(-W\rho^2 + \frac{a_k^2}{l^2}\frac{\mu_k^2}{\Xi_k}(r^2+a_k^2)\big)\frac{dt^2}{l^2}+\frac{r^2+a_k^2}{\Xi_k}\mu_k^2(d\phi_k^R)^2 \\ \nonumber
&+& 2 a_k \mu_k^2(r^2+a_k^2)\frac{dt d\phi_k^R}{l^3 \Xi_k}+\sum_{i \neq p \neq k}\frac{r^2 + a_i^2}{\Xi_i} \mu_i^2 d\phi_i^2 + r^2 \mu_p^2 d\phi_p^2.
\end{eqnarray}
Now, we introduce the following new azimuthal coordinate 
\begin{equation}\label{varphik}
\varphi_k=\frac{\phi_k^R}{\Xi_k}.
\end{equation}
After taking the limit $a_k \rightarrow l$, we have $\ev{\omega} \rightarrow \usev{\omega}$ and $d \ev{\gamma}^2 \rightarrow d\usev{\gamma}^2$  as
\begin{eqnarray}\label{omegas}\nonumber
\usev{\omega}&=&(\usev{W}+\mu_k^2)dt - l \mu_k^2 d\varphi_k - \sum_{i \neq j \neq k}\frac{a_i \mu_i^2 d\phi_i}{\Xi_i}, \\ \nonumber
d\usev{\gamma}^2&=&-\bigg((\usev{W} +\mu_k^2)\rho^2 + \mu_k^2l^2\bigg)\frac{dt^2}{l^2} +\rho^2 \mu_k^2 d \varphi^2\\ 
&+&\frac{2 \rho^2 \mu_k^2 dt d\varphi}{l} + \sum_{i \neq p \neq k}\frac{r^2 + a_i^2}{\Xi_i} \mu_i^2 d\phi_i^2 + r^2 \mu_p^2 d\phi_p^2.
\end{eqnarray}
Finally, by putting everything together, we construct the ultraspinning EVH solution geometry, given by
\begin{equation}\label{USEVHMetric2}
ds^2= d\usev{\gamma}^2 + \frac{2m}{\usev{U}} \usev{\omega}^2 + \frac{\usev{U}dr^2} {\usev{X}}+d \usev{\Omega}^2,
\end{equation}
where
\begin{eqnarray}
\usev{U}&=&\bigg(\rho^2 \mu_p^2+r^2\mu_k^2+r^2\sum_{i \neq p \neq k}\frac{\mu_i^2 \rho^2}{r^2+a_i^2 }\bigg)\prod_{j \neq p\neq k}^{N}(r^2+a_j^2) \nonumber \\
\usev{X}&=& \frac{\rho^4}{l^2}\prod_{i \neq p\neq k}^{N}(r ^2+a_i^2)-2m.
\end{eqnarray}
Also the EVH constraint Eq. \eqref{EVHlimit} at the ultraspinning limit leads to 
\begin{equation}\label{EVHUSC2}
m=\frac{l^2}{2} \prod_{i\neq p \neq k}^{N}a_i^2.
\end{equation}
Finally, to compactify the new coordinate $\varphi_k$, one can identify it such that $\varphi_k \sim \varphi_k + \mu$. Where $\mu$ is a dimensionless parameter. 

It should be noted that as is discussed in \cite{1504.07529}, we are only allowed to take the single ultraspinning limit. Trying to have a multi-ultraspinning EVH geometry by additional limit such as $a_j \rightarrow l$, leads us to a divergence in the $g_{\mu_j\mu_k}$ component. By which this divergency can not be removed by a new rescaled azimuthal coordinate. 

\subsection{ EVH limit of ultraspinning black holes}
Here, we explore the upper path in Fig. 1. We should start with the ultraspinning version, which is presented in Eq. \eqref{USMetric1}. Then its EVH limit can be studied by imposing the one-cycle EVH definition Eq. \eqref{EVH}.

Using $\us{X}(r_{+})=0$, that determines the horizon location, we simplify the horizon area of the US-KAdS metric given by relation Eq.  (\ref{Area-US}), as
\be
{A_{H}}=\frac{2 \bar{\mathcal{A}}_{d-2}\;m\; r_{+}}{1+\frac{r_{+}^2}{l^2}}\prod_{i\neq k}^{N}\frac{1}{\Xi_{k}}.
\ee
It suggests that the vanishing horizon limit at fixed $m$, only happens in the following circumstances
\begin{center}
 $r_{+} \to \ep$, \qquad \qquad $\frac{r_{+}^2}{l^2} \to \infty$, \qquad or \qquad $\Xi_{i} \to \infty\; (l \to 0)$.
\end{center}
Consistency with the horizon location at $\us{X}(r_{+})=0$ and extremality limit restrict above conditions to 
\begin{center}
 $r_{+}=l \sim \ep$, \qquad or  \qquad $r_{+} \sim \ep, a_{p \neq k} \sim \ep^2$.
\end{center}
The former condition may gives rise a non-one-cycle EVH family solution, that obey $A_H/T^5=finite$. However, this kind of limit does lead to an ill-defined geometry. The latter case leads to the following EVH condition
\be
m=\frac{l^2}{2}\prod_{i \neq p\neq k }^{N}a_{i}^2.
\ee  
This result is exactly equal to condition Eq. \eqref{EVHUSC2}, that we have already derived for ultraspinning limit of the EVH KAdS solution.

Furthermore, in the case of NEVH-US-KAdS, we find the new following limits
\be \label{NearEVHlimitUS}
r_{+} =  \r \, \ep,  \qquad  a_{p} =  \a \,\, \ep^2, \qquad m=\left(1+\m \, \ep^2\right)\frac{l^2}{2}\prod_{i \neq p \neq k }^{N}a_{i}^2  , \quad \ep \to 0,
\ee 
where
\begin{gather}\label{Delta_m_US}
\m=  \frac{\a^2}{\r^2}+\frac{\r^2}{\l3^{2}}, \qquad \l3^{-2}\equiv \frac{2}{l^2}+\sum_{i \neq p\neq k}^{N}a_{i}^{-2}.
\end{gather}
We also find temperature and area of the horizon at near-EVH limit Eq.  (\ref{NearEVHlimitUS}) as
\begin{eqnarray}\label{Thermo-nearEVHUS}
T=\frac{1}{4\pi}(\frac{1}{\l3^2}-\frac{\a^2}{\r^4})\,\r \ep, \qquad  \qquad A_{H}=l^2 \bar{\mathcal{A}}_{d-2}\left(\prod_{i \neq p \neq k}^{N}\frac{a_{i}^2}{\Xi_{i}}\right)\;  \r \ep\;,
\end{eqnarray}
where obviously shows that the ratio $A_{H}/T$ is finite for both near-EVH and EVH ($\ep \to 0$) limits.

We emphasize, these results can be derived by simply taking the ultraspinning limit $a_{k} \to l$ onto the (near) EVH conditions Eq. \eqref{NearEVHlimit}.

Finally by applying the EVH conditions \eqref{NearEVHlimit} onto \eqref{USMetric1}, one can straightforwardly construct the ultraspinning EVH solution, which the resulting geometry is exactly same as the metric \eqref{USEVHMetric2}. 
We therefore immediately conclude that the ultraspinning and EVH limits commute with each other ($p\neq k$). Namely, beginning with a general Kerr-AdS black hole and applying both ultraspinning and EVH limits in either order, the resulting geometry is perfectly the same. We call this geometry ``EVH-US-KAdS". In other words, taking limits either via the upper path or the lower one in Fig. \ref{Fig1} give us the exactly same metric.

Note, due to the EVH configuration Eq. \eqref{NearEVHlimitUS} that imply for vanishing one rotation parameter, thus it is not applicable to coincide the US and (near)EVH directions. Therefore, these two limits are independent.

It is worth mentioning that, our obtained US-EVH black hole Eq. \eqref{USEVHMetric2} seems to suffer some poles at $\mu_k=0$. But as we will show explicitly in appendix Eq. \eqref{App1}, these poles can be viewed as some punctures which are removed from the spacetime, indicating our new EVH black hole has a noncomapct manifold in $(t,r\neq r_+) = const.$ slices.
\section{Near horizon geometry}\label{SecNH}
It has been shown that the near horizon geometry of EVH solutions in Einstein-$\Lambda$ theory with negative cosmological constant enjoys an AdS$_3$ subspace in the near horizon \cite{1107.5705,1504.03607,1512.06186}. Since the ultraspinning KAdS black holes possess a particular noncompact horizon with two punctures, in the next we explicitly study the near horizon of US-EVH-KAdS black holes Eq. \eqref{USEVHMetric2}. We also show that they admit AdS$_{3}$ at EVH limit, and BTZ at the near-EVH limit. In addition, using the similar strategy that was taken in the previous section, we will demonstrate that for the EVH KAdS black holes the near horizon limit and the ultraspinning limit commute with each other.
\subsection{Near horizon of EVH-US-KAdS}
To obtain the near horizon geometry of the ultraspinning EVH black holes  Eq. \eqref{USEVHMetric2}, we use the following coordinate transformations
\begin{gather}\label{NearHorizonLimitUS}
r=\lambda \; \nh{r} , \qquad t= \frac{\nh{t}}{\lambda }, \qquad \phi_{p} = \frac{\nh{\phi}_p}{\lambda },\qquad  \varphi_{k }= \nh{\varphi}_{k}+\frac{\nh{t}}{l}, \qquad\; \phi_{i}= \nh{\phi}_{i}+\Omega_{\text{S}}^{i} \; \; (i \neq p \neq k),
\end{gather}
where we used $\r=0$, as horizon radius, and  $\Omega_S^{k}=\frac{1}{l}$, which denotes the angular velocity on the horizon at the ultraspinning direction $\varphi_k$. The all rest of angular velocities are given by Eq.  \eqref{Thermo-US}. Finally by taking the limit $\lambda \rightarrow 0$, we find the near horizon geometry as
\begin{gather}\label{NHG-USEVH}
ds^2=\mu_{p}^2 \; (-\frac{\nh{r}^2}{\l3^2}d\nh{t}^2+\frac{\l3^2}{\nh{r}^2}d\nh{r}^2+\nh{r}^2 d\nh{\phi}^2)+ds_{\mathcal{M}_{d-3}}^2,
\end{gather}
\begin{eqnarray}\label{Md-3}
ds_{\mathcal{M}_{d-3}}^2&=&
 2 \sum_{i\neq k \neq p }^{N}(\frac{ \mu_{k}^2\, \mu_{i}^{2}\,a_i\,l}{\mu_{p}^2 \, \Xi_i}d\nh{\phi}_i d\psi_k-\frac{\mu_{i}\, a_{i}^2}{\mu_{k}\, \Xi_{i}}d\mu_{i}\,d\mu_{k} )
+(\usev{W}+\mu_{k}^2)\frac{l^2}{\mu_{k}^2}\;d\mu_{k}^2\nonumber\\
&+&\sum_{i,j \neq p\neq k}{(h_{ij}d\mu_{i}d\mu_{j}+k_{ij}d\nh{\phi}_{i}d\nh{\phi}_{j})}+\frac{\mu_{k}^{4}\,l^2}{\mu_{p}^{2}}d\nh{\varphi}_{k}^2,
\end{eqnarray}
where 
\bea\label{metricH}
h_{ij}=\frac{a_i^2\,\mu_i^2}{\Xi_i}\, \delta_{ij}+\frac{\mu_i^2\, \mu_j^2}{\mu_p^2} \frac{a_i\, a_j }{\Xi_i\,\Xi_j}, \qquad
k_{ij}=\frac{a_i^2}{\Xi_i}\; \delta_{ij} \,.
\eea

It obviously contains an AdS$_3$ sector that arises from the vanishing of the rotation parameter $a_p$ corresponding to the $\phi_p$ direction. The rest part shows a $S^{d-3}$ manifold. It should be noted this part of the near horizon geometry inherits the non-compactness structure from the origin EVH-US solution. It can be viewed topologically as a sphere with two punctures. Furthermore, this geometry is regular everywhere except at $\mu_{p}=0$, where Kretschmann scalar diverges. Therefore, the US limit cannot resolve singularity of the near horizon of EVHs.

\subsection{Near horizon of near EVH-US-KAdS}
In this section we investigate a more general near horizon geometry for US-KAdS solution Eq.  (\ref{USEVHMetric2}) at near-EVH limit, via taking both near-EVH and near horizon limits together. By employing the following transformation
\begin{gather}\label{NearHorizonLimitUS2}
r-\r=\lambda \; \nh{r} , \quad t= \frac{\nh{t}}{\lambda }, \quad \phi_{p} = \frac{\nh{\phi}_p}{\lambda },\quad \phi_{i}= \nh{\phi}_{i}+\Omega_{\text{S}}^{i} \; t,\quad \varphi_{k }= \nh{\varphi}_{k}+\frac{t}{l}, \; \quad (i \neq p \neq k),
\end{gather}
along with the near-EVH limit Eq. \eqref{NearEVHlimitUS}, then taking the limits $\lambda \sim \ep \to 0$, we find the near horizon geometry as
\begin{gather}\label{NHG-USEVH2}
ds^2=\mu_{p}^2 \; \big{(}-f\;d\nh{t}^2+\frac{d\nh{r}^2}{f}+\nh{r}^2(d\nh{\phi}_{p}-\frac{\a}{\nh{r}^2}d\nh{t})^2\big{)}+ ds_{\mathcal{M}_{d-3}}^2,\\  f(\nh{r})=\frac{(\nh{r}^2-\r^2)(\nh{r}^2-\;\frac{\a^2\;\l3^2}{ \r^2})}{\l3^2 \; \nh{r}^2 },\nonumber
\end{gather}
where $ds_{\mathcal{M}_{d-3}}^2$ is given by Eq. \eqref{Md-3}. The first term represents a BTZ black hole. In the case of $\r=0=\a$ this metric reduces to Eq. \eqref{NHG-USEVH}. The constructed near horizon geometry for ultraspinning black holes in both EVH and near-EVH limits lead to a three-dimensional maximally symmetric subspace as AdS$_3$ and BTZ black holes respectively. Thus, these results obey the previous argument regarding EVH solutions. Also in the near horizon of EVH-US-KAdS, the $(d-3)$-dimensional subspace $ds_{\mathcal{M}_{d-3}}^2$ is a non-compact manifold.


\subsection{Near horizon of EVHs under US limit}
It was shown that for a large class of extremal black holes the near horizon and ultraspinning limits commute with each other  \cite{1512.07597,1611.02324,1702.03448}. Now, we explore whether the near horizon limit of an EVH solution commute with the ultraspinning limit. For this purpose, we follow two different paths as shown in Fig. \ref{Fig2} to get the near horizon of EVH-US-KAdS.
\begin{figure}[t]
\begin{center}
\includegraphics[scale=1]{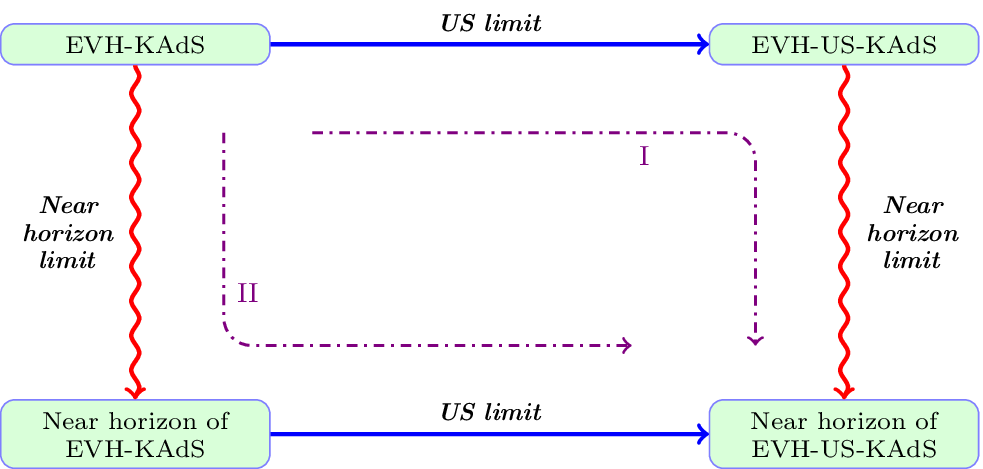}
\caption{This diagram shows two different orders of limit for an EVH Kerr-AdS black hole (EVH-KAdS). Horizontal arrows (blue) represent the ultraspinning (US) limit. The vertical ones (red) show the near horizon limit. We show that in both paths the resulting geometry (near horizon of EVH-US-KAdS) are perfectly equivalent.}\label{Fig3}
\end{center}
\end{figure}
We have already obtained the near horizon of EVH-US-KAdS Eq. \eqref{NHG-USEVH} by taking the near horizon limit of a given EVH-US-KAdS, (path (I) in Fig. \ref{Fig2}). Now, we show one can also get the same geometry by starting from the near horizon of EVH-KAdS Eq. \eqref{NearHorizonNearEVH} and then taking the ultraspinning limit over it (lower path (II) of Fig. \ref{Fig2}). 

Let us follow the lower path of Fig.\ref{Fig2}.
In order to perform the ultraspinning limit along $\nh{\phi}_{k}$ direction, we just need to rescale this coordinate as $\nh{\varphi}=\nh{\phi}_{k}/\Xi_{k}$, then we take the $a_{k} \to l$ limit. Finally, we find the metric Eq. \eqref{NHG-USEVH}.

Also, we require to compactify the new coordinate $\varphi_k$ like Eq.  (\ref{mu}), as $\varphi_k \sim \varphi_k +\mu$. It is worthwhile to mention, since  the near horizon geometry is relaxed from the asymptotic structure, it is not necessary to use the ARF coordinate to apply the ultraspinning limit. Based on the above arguments, we conclude the near horizon and ultraspinning limits commute with each other for the EVH-KAdS black holes.

One can show that the near horizon of (N)EVH-US-KAdS admits $SO(2,2)\times U(1)^{\left[\frac{d-1}{2}\right]}$ isometry. The most general five-dimensional solutions with $SO(2,2)\times U(1)^2$ isometry has been proposed in \cite{1409.1635}, which convenience for both positive and negative cosmological constant $\Lambda=\pm 6/l^2$, is given by
\be\label{SO22-5d}
ds^2=\frac{a^2}{(1-\frac{\Lambda a^2}{6})^2}\cos^2\theta \, ds_3^2+\frac{a^2\,\cos^2\theta}{\Delta(\theta)}d\theta^2+\frac{a^2}{(1+\frac{\Lambda a^2}{6})^2}\Delta(\theta)\tan^2\theta\,d\phi^2
\ee	

In fact, this generic metric can be considered as a near horizon geometry of a five-dimensional EVH black hole solution in the same theory, in which the two-dimensional part $(\theta,\phi)$ is a compact manifold.

Now, in the case of negative cosmological constant and by applying the ultraspinning limit on above metric, we can straightforwardly reproduce the near horizon geometry of our EVHUS-KAdS black holes Eq.  (\ref{NHG-USEVH2}) in five dimensions as 
\be\label{USSO22-five-dimensional}
ds^2=\frac{l^2}{4}\cos^2\theta \, ds_3^2+\frac{l^2\,\cos^2\theta}{\sin^2\theta}d\theta^2+l^2\frac{\sin^2\theta}{\cos^2\theta}\,d\varphi^2
\ee		
where $\varphi$ is ultraspinning azimuthal coordinate by periodicity $\varphi\sim\varphi+\mu$. 

We note that, this metric describes a five dimensional $SO(2, 2)$ invariant solution to Einstein-$\Lambda$ $(\Lambda<0)$ theory. As we will discuss in Appendix \ref{App1}, this geometry has two punctures at $\theta=0, \pi$, providing a non-compact manifold, but finite area for all $(t, r\neq_+)=const.$ slices. 

\section{Thermodynamics and EVH/CFT}\label{SecThermo}
As we have shown in the previous section, the near horizon geometry of EVH-US-KAdS is as warped product of a locally AdS$_3$ subspace with a $\mathcal{M}_{d-3}$ Eq. \eqref{NHG-USEVH}. In the current section, we will first study the thermodynamic of three-dimensional near horizon geometry Eq. \eqref{NHG-USEVH2} by comparing them with the thermodynamic quantities of origin EVH-US-KAdS black hole Eq. \eqref{USEVHMetric2}.
We also briefly study the CFT$_2$ dual description of the near horizon geometry.

\subsection{Thermodynamics of 3-dim vs. $d$-dim }
Now, we explore a relation between thermodynamics quantities of this three-dimensional part with their $d$-dimensional origin black hole. By performing a standard compactification over $\mathcal{M}_{d-3}$, we find a three-dimensional gravity that governs dynamics of the AdS$_3$ part (see \eg \cite{1301.3387}). Standard reduction for Hilbert-Einstein action over the metric anstaz Eq. \eqref{NHG-USEVH} shows a simple relation between three dimensional Newton constant and its $d$-dimensional cousin, as (see \eg \cite{1301.3387})
\be\label{G3}
G_{3}=\frac{2\pi G}{l^2 \tilde{\mathcal{A}}_{d-2}}\prod_{i \neq p \neq k}^{N}\frac{\Xi_{i}}{a_{i}^2}.
\ee
In the case of NEVH black hole solution, the AdS$_3$ sector is actually a pinching BTZ. Then, for completeness, we compare its thermodynamics with the NEVH-US-KAdS ones.  Thermodynamic quantities of the BTZ black hole Eq.  (\ref{NHG-USEVH}) is given by
\be\label{BTZ-Thermo}
\begin{gathered}
\btz{M}=\frac{\a^2 \l3^2+\r^4}{8 G_{3} \l3^{2} \r^2} \;\epsilon\;,  \ \ \ \ \btz{S}=\frac{\pi \r}{2 G_{3}}\; \epsilon\;,  \ \ \ \ \btz{J}= \frac{\a}{4 G_{3}} \; \epsilon\;,  \\
\btz{T}=\frac{1}{2 \pi \l3^2}\frac{\r^4-\a^2 \l3^2}{\r^3}\;,\ \ \ \ \btz{\Omega}=\frac{\a}{\r^2}\;,
\end{gathered}
\ee
where $\epsilon$ comes from integration over the pinching azimuthal coordinate direction $\phi_p$ with $2\pi \ep $ periodicity. On the other hand, thermodynamic quantities for NEVH-US-KAdS metric can be straightforwardly obtained by applying the near-EVH limit over US-KAdS thermodynamic quantities Eq.  \eqref{Thermo-US}. Then by comparing them with the corresponding quantities of BTZ Eq.  (\ref{BTZ-Thermo}), we have
\be\label{NearHorizonQuantities}
\begin{aligned}
T&=&\epsilon \,\btz{T} \;, \qquad \Omega^{p}&=&\btz{\Omega}\;, \\ 
S&=&\btz{S}\; , \qquad J_{p}&=&\epsilon\, \btz{J}\;.
\end{aligned}
\ee
Furthermore, the mass of the BTZ black hole is proportional to the mass parameter deviation $\m$ above the EVH surface
\be
\btz{M}=\frac{\m}{8 G_{3}}.
\ee
Following \cite{1107.5705,1301.3387,1305.3157} we can check that the first law of thermodynamics for the BTZ black hole is identical to the first law for NEVH-US-KAdS, as far as parameter variations normal to EVH surface are concerned. To show this issue, we expand thermodynamics quantities around the EVH surface $\ep=0$
\be
\begin{aligned}
T&= T_{(1)}\; \ep +\O(\ep^3), &   S\;&= S_{(1)} \;\ep +\O(\ep^3), \qquad M=M_{(0)}+M_{(2)}\ep^2+\O(\ep^4),
\\
J_{i}&=J_{i\;(0)}+J_{i\;(2)}\ep^2+ \O(\ep^4), & \Omega_{i}&=\Omega_{i\;(0)}+\Omega_{i\;(2)}\ep^2+ \O(\ep^4),\quad i\neq  p,
\\
J_{p}&=J_{p\;(2)}\ep^2+ \O(\ep^4), & \Omega_{p}&=\Omega_{p\;(0)}+\Omega_{p\;(2)}\ep^2+ \O(\ep^4).
\end{aligned}
\ee

Now the expansion of the first law up to $\O(\ep^2)$ gives
\begin{align*}\label{firstlaw}
T_{}\delta S_{}&=\delta M-\sum_{i}\Omega_{i}\delta J_{i},\\
T_{(1)}\delta S_{(1)}\ep^2&=\left(\delta M_{(2)}-\sum_{i\neq p}(\Omega_{i \;(0)}\delta J_{i\;(2)}+\Omega_{i \;(2)}\delta J_{i\;(0)})-\Omega_{p\;(0)}\delta J_{p \; (2)}\right)\ep^2\\
&+\left(\delta M_{(0)}-\sum_{i}\Omega_{i\;(0)}\delta J_{i \;(0)}\right)+\O(\ep^4).
\end{align*}
Zeroth order expressions represent the variations along the EVH surface, that one can consider them as the EVH first law thermodynamics. While, second order terms may express variations normal to the EVH surface, that they can change the entropy and temperature of the EVH solution. However, it is easy to see that the expressions $\Omega_{i \;(2)} \delta J_{i \;(0)} \propto \delta a_{i}$ would not be able to produce normal variations. So, we set them equal to zero. Therefore, only the variations normal to the EVH surface are taken into account. Now, by noting relations Eq.  \eqref{NearHorizonQuantities} we find the following explanations as thermodynamics first law for normal variation ($\delta_{\perp}$)  (up to $\O(\ep^2)$) 
\be
\begin{aligned}
(T_{\text{BTZ}}\delta S_{\text{BTZ}}+ \Omega_{\text{BTZ}} \delta J_{\text{BTZ}}) \;\ep&=(T_{(1)}\delta_{\perp} S_{(1)} +\Omega_{p\;(0)}\delta_{\perp} J_{p \; (2)})\ep^2\\
\delta M_{\text{BTZ}} \; \ep
&=( \delta_{\perp} M_{(2)}-\sum_{i\neq  p}\Omega_{i \;(0)}\delta_{\perp} J_{i\;(2)})\; \ep^2.
\end{aligned}
\ee
The above last equality suggests the following relation between the masses of the BTZ black hole and the near EVH black hole
\be
M_{\text{BTZ}} \; \ep = M_{(2)} \ep^2-\sum_{i\neq  p}\Omega_{i \;(0)} J_{i\;(2)}\ep^2=\frac{\m}{8 G3}\ep^2.
\ee
This argument shows excitations normal to the EVH surface change, conserve charges of the EVH-US-KAdS such that they satisfies a tree-dimensional first law of thermodynamics. In addition, it is easy to check that the mentioned second order quantities satisfy the following relation
\be
\left(\frac{l^2 \tilde{\mathcal{A}}_{d-2}}{16\pi G_d}\prod_{i \neq p \neq k}^{N}\frac{a_{i}^2}{\Xi_{i}}\right) \m=(\frac{1}{2}T_{(1)}S_{(1)}+\Omega_{p\;(0)}J_{p\;(2)} ),
\ee
 that is same as the Smarr formula of a BTZ black hole. This suggests that the excitation of EVH may be described by a three-dimensional gravity. Since this three-dimensional gravity lives in an AdS$_{3}$, it may have a CFT$_2$ dual description \cite{1107.5705}.

\subsection{EVH/CFT correspondence}
Appearing of an AdS$_{3}$ factor in the near horizon of EVH black holes, and  the result that the entropy of original near-EVH black hole is equal to the entropy of the near horizon geometry that is a BTZ black hole, suggest a CFT$_2$ dual description for low energy excitation of EVH black holes. \footnote{This idea may be supported by checking this fact that near horizon limit is  a decoupling limit indeed. This point has been shown for rotating $4$d Kaluza-Klein EVH black holes \cite{1107.5705}.} These observations inspired the
authors of \cite{1107.5705} to propose the EVH/CFT correspondence analogues to familiar Kerr/CFT correspondence. That conjectures for any extreme Kerr black hole one can find a $2$d dual conformal field theory \cite{0809.4266}.
This idea has been fully studied for large families of black holes \cite{0811.4393}, indicating the microscopic entropy of the dual CFT via the Cardy formula exactly agree with the Bekenstein-Hawking entropy of the extreme kerr black hole. Also the validity of this correspondence has been confirmed for ultraspinning non-compactness horizon black holes in \cite{1512.07597, 1611.02324, 1702.03448}. 

The basic idea of EVH/CFT is that the low energy excitations of an EVH geometry attributed to a CFT$_2$, that lives on the boundary of a pinching AdS$_ {3} $. The central charge of CFT$_2$ can be derived by following the primary example of Brown and Henneaux \cite{Brown:1986nw}, that by imposing a set of consistent boundary conditions, the asymptotic symmetry group of AdS$_3$ is enhanced to two copies of the Virasoro algebra.

Here, we establish the ultraspinning EVH solution Eq.  (\ref{USEVHMetric2}) exhibit a well-defined EVH/CFT by following \cite{1107.5705}. Central charges of CFT$_{2}$ dual to the pinching AdS$_3$ read as  
\be
c=\frac{3 \, \l3 \, \ep}{2 G_{3}}=\frac{3\;l^2 \l3  \bar{\mathcal{A}}_{d-2}}{4\pi G}\prod_{i \neq p \neq k}^{N}\frac{a_{i}^2}{\Xi_{i}} \;\ep.
\ee

On the other hand, the Virasoro generators $L_0$ and $\bar{L}_0$ can be expanded in terms of conserved charges of the BTZ black hole using the standard method
\be
\begin{aligned}
L_0-\frac{c}{24}&=& \frac{1}{2}(\l3\,M_{\text{BTZ}}+J_{\text{BTZ}}), \\
\bar{L}_0+\frac{c}{24}&=& \frac{1}{2}(\l3\,M_{\text{BTZ}}-J_{\text{BTZ}}),
\end{aligned}
\ee

Now, using the Cardy's formula \cite{Cardy:1986ie} which counts microstates of an unitary and modular invariance two-dimensional CFT, we compute the entropy of  CFT$_2$ as
\bea
S&=&2\pi \sqrt{\frac{c}{6}(L_{0}-\frac{c}{24})}+2\pi \sqrt{\frac{c}{6}(\bar{L_{0}}-\frac{c}{24})}\nonumber \\
&=& \frac{\pi \r}{2 G_{3}} \; \ep=\frac{l^2 \bar{\mathcal{A}}_{d-2}}{4 \; G}\left(\prod_{i \neq p \neq k}^{N}\frac{a_{i}^2}{\Xi_{i}}\right)\;  \r \ep.\;
\eea
This microscopic entropy is exactly equivalent with the entropy of the BTZ black hole Eq.  \eqref{BTZ-Thermo} and the entropy of the original NEVH-US black hole  Eq.  \eqref{Thermo-nearEVHUS}. This result confirms the validity of the EVH/CFT proposal for our non-compact ultraspinning EVH black hole solution. 

Note, the rotation parameters $a_i$'s and the dimensionless parameter $\mu$ Eq.  \eqref{Amu} appear in the central charge of the dual CFT$_2$. So, by changing their values, one can move through the various CFTs. Further study of this point could be interesting.
\subsection*{Finite entropy in vanishing horizon limit}
It is possible to keep entropy finite in the EVH limit while the horizon area goes to zero. To do this issue, following the \cite{1107.5705} argument, the Newton constant $G$ should scale the same as the horizon area and temperature. It means we need to promote the EVH to the following triple scaling \cite{1011.1897,1512.06186}
\be
A_{H} \sim \ep,\quad T \sim \ep, \quad G \sim \ep, \quad \ep \rightarrow 0.
\ee
Clearly this scaling  also implies a finite central charge $c$. However, instead of scaling $G$, there is another option in the case of EVH-US-KAdS. Thanks to Eq.  \eqref{mu} for our ultraspinning black hole, one can assume 
\be
\mu \, \ep =  finite \; \; \;\Longrightarrow \; \; \; \bar{\mathcal{A}} \sim \ep, \quad \ep \to 0.
\ee
Although this scaling does not change the d-dimensional Newton constant $G$, it implies $G_{3} \sim \ep$ that leads to a finite value for central charge $c$.
\subsection{Ten-dimensional embedding and AdS$_5$/CFT$_4$}
It was shown in \cite{Cvetic:1999xp} that all solutions to \eqref{Metric1} can be uplifted to ten-dimensional type IIB supergravity solutions. Using the ansatz \cite{Cvetic:1999xp} the ten-dimensional embedding of  five-dimensional US version \eqref{USMetric1} to on-shell ten-dimensional supergravity has the following description
\be
ds^2_{10}=ds^2_{5}+l^2\sum_{i=1}^{3}(d\bar{\mu}_i^2+\bar{\mu}_i^2d\psi_i^2)
\ee
where $ds^2_5$ is given by the metric \eqref{USMetric1} in five dimensions with rotation parameters $a$ and $b$, which the former would be the US direction. Also  $\bar{\mu}_i$ are functions defining an unit $2$-sphere
\be\label{uplifft}
\bar{\mu}_1=\sin \alpha \cos\beta, \qquad  \bar{\mu}_2=\sin \alpha \sin\beta, \qquad \bar{\mu}_3=\sin\beta,
\ee
Using AdS/CFT, the above ten-dimensional black holes solution corresponds to a four-dimensional $\mathcal{N}=4$ SYM with conformal dimension $\Delta$, and $SO(4)$ spins $\mathcal{S}_{a}$ and $\mathcal{S}_{b}$ as follow
\be\label{SYMUS}
{\Delta=l\,M_{us}=\frac{\mu\, m (2+\Xi_b)}{2\pi l^3 \Xi_b}N^2,\qquad \mathcal{S}_a=J_a=\frac{\mu\, m }{2\pi l^2 \Xi_b}N^2,  \qquad  \qquad \mathcal{S}_b=J_{b}=\frac{\mu \,b\,m}{2\pi l^3 \Xi_b^2}N^2,}
\ee
where we used the standard relation $G_{10}=G_5 \,\pi^3\,l^5=\frac{\pi^4 l^8}{2 N^2}$ between five and ten dimensional Newton's constant. Also $N$ is the RR five-form flux over $S^5$ with the radius $l$, defining as $l^4=4\pi g_s l_s^4\,N$. 

Analogues, the embedding of EVH black hole \eqref{EVHMetric1} to ten-dimensional type IIB supergravity solutions has the same explanation with \eqref{uplifft}, except the $ds^2_5$ is replaced with the \eqref{EVHMetric1} in five dimension, and we choose the rotation parameter $b$ as the EVH direction. Its dual $\mathcal{N}=4$ SYM carries following charges 
\be\label{SYMEVH}
{\Delta=l\,M_{EVH}=\frac{a^2 }{2l^3\,\Xi_a^2}N^2,\qquad \mathcal{S}_a=J_a=\frac{a^2 }{2l^3 \Xi_a}N^2,  \qquad  \qquad \mathcal{S}_b=0,}
\ee
Furthermore, the ten-dimensional embedding of our obtaining EVH-US-KAdS black holes \eqref{USEVHMetric2} corresponds to a four-dimensional CFT  with the following charges
\be\label{SYMUSEVH}
{\Delta=\frac{\mu}{2 \pi\,l}N^2,\qquad \qquad \mathcal{S}_a=\frac{\mu }{4 \pi}N^2,  \qquad  \qquad \mathcal{S}_b=0.}
\ee
As we see, all above charges corresponding to three different classes of solutions are scaled like $N^2$. Once notes, in the cases of the US and EVH-US solutions the  thermodynamic parameter $\mu$ contributes to the CFT charges. 
\section{Super-entropic NEVH black holes}\label{SecSuper}
It was shown in \cite{1411.4309,1504.07529} that this class of ultraspinning black holes exhibits a distinguished property in the context of extended phase space thermodynamics, where the variation of the cosmological constant contributes in the first law of black hole thermodynamics. These ultraspinning black holes violate the conjecture ``Reverse Isoperimetric Inequality" as the first counter example.

Here, we shall investigate whether  the US-EVH-KAdS black hole Eq.  (\ref{USEVHMetric2}) is a super-entropic EVH black hole in an extended phase space background. Since, EVHs have a vanishing horizon area (entropy), it seems naively they never provide a super-entropic black hole. However, it is intriguing to carefully analyze the isoperimetric inequality for obtaining our (near) EVH ultraspinning black holes.

In a theory of gravity that the cosmological constant arises as the vacuum expectation value of a field, one can consider its variation in the first law of thermodynamics. It was argued in \cite{0904.2765}  that the thermodynamic conjugate of $\Lambda$ which is a finite and negative quantity $\Theta=8\pi G(\partial M/\partial \Lambda)$, contributes in the extended first law as $\Theta\, d\Lambda$.

If we assume the cosmological constant as a thermodynamic pressure 
\be
P=-\frac{1}{8\pi G}\Lambda= \frac{(d-2)(d-1)}{16 \pi G l^2},
\ee
then the new term $\Theta$ looks like $V\delta P$. The effective volume $V$ can be interpreted as thermodynamic conjugates of pressure $P$. Thus, the extended first law of black hole thermodynamics takes following explanation by a geometric argument \cite{0904.2765} 
\be\label{firstLaw}
\delta M=T\;\delta S+\sum_{i}^{N}\Omega^{S}_{i}\; \delta J_{i}+V\; \delta P,
\ee
where $M$ should be considered as the  \emph{enthalpy}  of spacetime rather than as the total energy of the system. Also the corresponding Samrr formula can be derived using a scaling argument based on Euler's theorem \cite{0904.2765}
\be\label{Smarr}
\frac{d-3}{d-2}M=TS+\sum_{i}\Omega_iJ_i-\frac{2}{d-2}VP,
\ee
Using this fact, the thermodynamic volume simply is given by $V=\left(\frac{\partial M}{\partial P}\right)_{S,J_{i}}$. The effective finite volume $V$ for a large class of black hole solutions were calculated in \cite{1012.2888}. For general multi-spinning spinning Kerr-AdS black hole Eq.  (\ref{Metric1}), the thermodynamic volume $V$ is given by
\be
\begin{aligned}
	V &=\frac{r_{+}A_{H}}{d-1}+\frac{8 \pi G}{(d-1)(d-2)}\sum_{i=1}^{N}a_{i}J_{i}.
\end{aligned}
\ee
In the case of static black holes it reduces to the simple relation $\frac{r_{+}A_{H}}{d-1}$, that naively matches with the geometric volume of the AdS-Schwarzschild black hole interior.

Let us consider the following known isoperimetric relation for a given volume $V$ and area $A$ in $(d-1)$-dimensions 
\be \label{R}
\mathcal{R}\equiv \left(\frac{(d-1)V}{\mathcal{A}_{d-2}} \right)^{\frac{1}{d-1}}\left(\frac{\mathcal{A}_{d-2}}{A}\right)^{\frac{1}{d-2}}.
\ee
For a connected region in Euclidean space, the isoperimetric relation  always satisfies  $\mathcal{R} \leq 1$, and equality occurs if and only if the region is a round ball. While, in the context of black hole physics there is a conjecture that states for horizon area $A$ and thermodynamic volume $V$, the isoperimetric inequality violates, namely $\mathcal{R} \geq 1$,  and AdS-Schwarzschild black hole saturates this bound \cite{1012.2888}. The first family of black holes that violate the \emph{reverse} isoperimetric inequality is addressed to ultraspinning KAdS presented in\cite{1411.4309,1504.07529}. Therefore, these new ultraspinning black holes are dubbed as super-entropic, because their entropy exceeds the expected maximal entropy corresponding to AdS-Schwarzschild black hole.

Now we analyze the isoperimetric inequality for our obtained US-NEVH black holes Eq. \eqref{USEVHMetric2} for various dimensions. We will show that only five-dimensional case violates the \emph{reverse} isoperimetric inequality.

\subsection*{NEVH-KAdS}
Thermodynamic volume of a near EVH-KAdS black hole reads
\be
\ev{V}=\frac{8 \pi G}{(d-1)(d-2)}\sum_{i\neq p}a_{i}\ev{J}_{i}+\frac{\mathcal{A}_{d-2}}{d-1}\left(\prod_{i \neq p }^{N}\frac{a_{i}^2}{\Xi_{i}}\right)\;  \r^2 \ep^2+\mathcal{O}(\ep^4),
\ee
where $\ev{J}_{i}$ is given by angular momentum $J_i$ Eq.  \eqref{Charges} at the EVH limit. Using the above relation and the near EVH horizon area Eq.  \eqref{TAevh}, we obtain the isoperimetric $\mathcal{R}$ for near-EVH KAdS black holes as
\be
\ev{\mathcal{R}}= \left(\frac{8 \pi G}{(d-2)\mathcal{A}_{d-2}}\sum_{i\neq p}a_{i}\ev{J}_{i}\right)^{\frac{1}{d-1}}\left( \prod_{i \neq p }^{N}\frac{a_{i}^2}{\Xi_{i}}\;\frac{1}{\ep \; \r}\right)^{\frac{1}{d-2}}+\mathcal{O}(\ep^\frac{d-1}{d-2}).
\ee
Obviously, it goes to infinity in the $\ep \to 0$ limit. This result confirms the EVH-KAdS black holes have the minimum entropy among the KAdS's, indicating that they obey the reverse isoperimetric inequality, as like as the other generic black holes.

\subsection*{NEVH-US-KAdS}
Now we explore the isoperimetric inequality for the near EVH-US-KAdS metric that its thermodynamic volume is given by
\be
\usev{V}=\frac{8 \pi G}{(d-1)(d-2)} \sum_{i\neq p\neq k}a_{i}{\usev{J}}+\frac{l^2\bar{\mathcal{A}}_{d-2}}{d-1}\left(\prod_{i \neq p \neq k }^{N}\frac{a_{i}^2}{\Xi_{i}}\right)\;  \r^2 \ep^2+\mathcal{O}(\ep^4),
\ee
where $\usev{J}$ denotes angular momentum Eq. \eqref{Thermo-US} at the EVH limit Eq.  \eqref{NearEVHlimitUS}. We note, in five-dimensional case, $\usev{V}\sim \ep^2$, and its volume vanishes at the EVH limit ($\ep \to 0$). Now using Eq.  \eqref{Thermo-nearEVHUS}, we find the isoperimetric ratio as
\be
\begin{gathered}
\ev{\mathcal{R}}=(\frac{\r}{l} \ep)^{\frac{1}{6}}+\mathcal{O}(\ep^\frac{7}{6}), \qquad \qquad d=5\\
\ev{\mathcal{R}}= \left(\frac{8 \pi G}{(d-2)\mathcal{A}_{d-2}}\sum_{i\neq p \neq k}a_{i}\ev{J}_{i}\right)^{\frac{1}{d-1}}\left( \prod_{i \neq p \neq k }^{N}\frac{a_{i}^2}{\Xi_{i}}\;\frac{l^2}{\ep \; \r}\right)^{\frac{1}{d-2}}+\mathcal{O}(\ep^\frac{d-1}{d-2}), \quad d> 5.
\end{gathered}
\ee
Now, upon taking the limit $\ep \to 0$, we find that in five-dimensions, the isoperimetric ratio for EVH-US-KAdS metric Eq.  (\ref{USEVHMetric2}) goes to zero. While it goes to infinity for higher dimensional cases ($d \geq 5$). In this sense, one can call five-dimensional NEVH-US-KAdS black hole as the first example of (near) EVH super-entropic black hole.

\section{Conclusion}\label{SecConclusion}
We have utilized the ultraspinning limit, introduced in \cite{1411.4309,1504.07529}, to construct a novel class of EVH black hole solutions. We have demonstrated that the ultraspinning general Kerr-AdS black holes, despite the non-compactness of their horizons, have a well-defined EVH configuration in odd-dimensional spacetime. Also, we have shown that the resulting EVH-US black holes can be exactly reproduced in a different way. Namely, by applying ultraspinning limit onto the EVH Kerr-AdS black holes. In fact, the ultraspinning and EVH limits commute with each other. Moreover, it is impossible to apply EVH and ultraspinning limits in the same direction, because it turns out to an ill-defined geometry. We have also discussed the obtained black holes present a non-compact manifold, but with finite area in all $(t,r \neq r_+) = const.$ slices.

It has been proved in \cite{1504.03607,1512.06186}, the EVH black holes in Einstein theory in the presence of a negative cosmological constant enjoy an AdS$_3$ subspace in their near horizon, where it would be replaced with a BTZ black hole in the near EVH cases. We have confirmed the validity of this statement for our EVH-US-KAdS black holes, that have
non-compact horizons. Additionally, we precisely investigated the near horizon limit of EVH-US-KAdS can be obtained by employing an appropriate US limit to the near horizon
geometry of a given EVH-KAdS black hole. Namely, as we have 
demonstrated in Fig. \ref{Fig3}, the ultraspinning limit commutes with the near horizon limit for EVH black holes.

We compared thermodynamics of NEVH-US-KAdS black hole with thermodynamics of a pinching BTZ. The result shows entropy, temperature, angular momentum and angular velocity along the EVH direction $\phi_{p}$ are the same, up to a $\ep$ factor. While, the mass of the BTZ black hole is proportional to changes of EVH-US-KAdS mass parameter above the EVH point. 

Also, by assuming the first law of thermodynamics for the NEVH-US-KAdS, and expanding it around the EVH point, we have reproduced the first law for a BTZ black hole when charge variations normal to EVH surface are concerned. In addition, we checked that the Smarr mass formula for the BTZ black hole is the same as the relation among the mass parameter variation above the EVH point, entropy, temperature, angular momentum and velocity along the EVH direction (pinching direction). 

One may note, in $d>5$ , Newton constant $G_{3}$ Eq.  \eqref{G3} and AdS$_3$ radius $\l3$ Eq.  \eqref{Delta_m_US} are given in terms of $a_{i\neq p \neq k}$'s.\footnote{In $d=5$, $G_3$ and $\l3$ are constant.} So, it is natural to take their variations into account when we deal with the first law of thermodynamics for the BTZ black hole. Naively, one may expect it should compensate  $(\Omega_{i \;(2)} \delta J_{i \;(0)})$  terms that are not zero for variations along the EVH surface. However, by this strategy, the first law is no longer valid by usual assumption about the pressure $P_{BTZ} = \frac{1}{8 \pi G_{3} \l3}$  and the thermodynamic volume $V_{\text{BTZ}}=\pi \r^2$. Although only in $d=7$ (where there is only one free $a_{i \neq p \neq k}$) it is possible to define a new effective volume to take care effects of $\Omega_{i \;(2)} \delta J_{i \;(0)}$  via $V_{eff} \delta P_{BTZ}$. These difficulties should be due to that in the our case $\delta G_{3} \neq 0$. It would be intriguing to study this problem further.

We have also briefly investigated the EVH/CFT proposal and AdS$_3$/CFT$_2$ correspondence for our new EVH-US-KAdS solutions. We then found the microscopic entropy of the dual CFT using Cardy formula, that it is exactly equal with the entropy of the origin NEVH black hole and its near horizon geometry. We then argued that both side entropies and central charge of the CFT dual would be finite even at the EVH point, as long as we keep finite the product of near EVH parameter $(\ep)$ to the periodicity of US limit direction $(\mu)$, at EVH limit $\ep \to 0$.

A generic KAdS black hole has a dual ultraviolet (UV) CFT which lives on a compact boundary $\mathbb{R} \times S^{d-1}$. Extremal KAdS has an AdS$_{2}$ throat in the near horizon which suggests an infrared (IR) CFT dual. Generically, compactness of the manifold leads to a gapped  spectrum of the CFT. It implies fragmentation problem and the impossibility of adding low energy excitations to the IR CFT \cite{hep-th/9812073}. This problem is improved for extremal planar black holes whose boundary is non-compact and an infinite volume factor $\mathbb{R}^{d-1}$  exists in their near horizon \cite{0907.2694}. \footnote{See \cite{1112.4664} for further discussion.}

Interestingly, US limit yields a non-compact boundary with a free parameter $\mu$  that allows having an infinite volume in the near horizon of extremal KAdS. Then, the dual CFT may admit a continuous spectrum and non-trivial physical excitations at IR CFT.  It would be interesting to study this issue further by investigating the relation between UV and IR Green's functions in (near) extremal limit of USKAdS, as well as their (near) EVH limit. \footnote{A similar research has been done for planar extremal black holes \cite{0907.2694}.} 

In order to find a super-entropic near EVH black hole, we also investigated the reverse isoperimetric inequality for NEVH-US-KAdS solution. Since the black hole horizon vanishes in the EVH limit. So, one may naively except any NEVH black hole should satisfy this inequality. However, a careful analyze shows that five-dimensional NEVH-US-KAdS in the US limit violates this inequality. In this sense, these  black holes near the EVH point are super-entropic. While in other dimensions we cannot find EVH super-entropic black holes.

We expect that the EVH limit of  recently obtained charged ultraspinning black holes in diverse dimensions \cite{1611.02324,1702.03448} can be established through an analogous strategy that we have presented.

A further study would be using hyperboloid membrane as another ultraspinning limit \cite{0806.1954,1206.4351} to find EVH hyperboloid black hole solutions. the combination of this limit with US limit (super-entropic limit) recently was studied by \cite{1512.02293}, demonstrating one can find multi-ultraspinning black holes. Also, both limits commute with each other and the resulting geometry has a particular structure. However, their thermodynamics is not fully understood, one can employ these various ultraspinning limits beside the EVH limit to generate new exact black hole solutions with a novel horizon topology and asymptotic structure.

Recently, some ideas about the holographic implication of the varying $\Lambda$ have been mentioned \cite{1404.5982,1406.7267,1608.06147,1610.02038}. It would be interesting to study the role of this variation in the context of the near horizon BTZ black hole and EVH/CFT. 


\section*{Acknowledgments}
 We would like to thank M. M. Sheikh-Jabbari for suggesting this project and numerous helpful comments and discussions. Also we thank K.Hajian, S. Sadeghian and H. Soltanpanahi for useful discussions. S.M.N. would like to thank the Institute for Research in Fundamental Sciences (IPM) for hospitality while this project was accomplished. M.H.V. thanks the hospitality of ICTP, Trieste, while this project completed and his work has been supported as Simons Visitor at the ICTP through the Simons Award of M. M. Sheikh-Jabbari.

\appendix

\section{The topology of $(t,r \neq r_{+})=const.$ slices}\label{App1}
Let us more study the geometric structure of obtained EVH-US-KAdS black hole Eq.  \eqref{USEVHMetric2}. We consider the induced metric on the geometry of constant $(t, r\neq r_+)$ surfaces, which has the following description
\begin{equation}\label{HorizonUSEVH}
ds^2_h= \rho^2 \mu_k^2 d \varphi^2
+ \sum_{i \neq p \neq k}\frac{r^2 + a_i^2}{\Xi_i} \mu_i^2 d\phi_i^2 + r^2 \mu_p^2 d\phi_p^2 + \frac{2m}{\usev{U}} \big(l \mu_k^2 d\varphi_k + \sum_{i \neq p \neq k}\frac{a_i \mu_i^2 d\phi_i}{\Xi_i}\big)^2 +d\usev{\Omega}^2,
\end{equation}
Let us focus on $d\tilde{\Omega}^2_{us}$ part Eq.  (\ref{OmegaUSEVH}). It seems there are some singularities at poles $\mu_k=0$. But by examining above induced metric near $\mu_k=0$, following the procedure used in \cite{1702.03448}, we show these poles are not part of the spacetime. 

For simplicity we restricted ourself to $\phi_{i \neq k }=const.$ and $\mu_{i\neq k}=const.$ slices. Now we expand the metric Eq.  (\ref{HorizonUSEVH}) in terms of $\mu_k$. In the leading order of limit $\mu_k \rightarrow 0$, we find the following explanation 
\begin{equation}\label{Horizon-k}
ds^2_h\approx\rho^2 \tilde{W}\bigg[\frac{d\mu_k^2}{\mu_k^2}+\frac{2 \,m}{\tilde{W}\rho^2\tilde{U} } l^2 \mu_k^2 d\varphi_k^2  \bigg].
\end{equation}

This metric exhibit a manifold with constant negative curvature on a quotient of the hyperbolic space $\mathbb{H}^2$. Indeed, the poles $\mu_k=0$, can be viewed as some sort of punctures that are removed from the spacetime. Namely, the $(t,r) = const.$ slices have no singularities at these poles. Therefore the obtained ultraspinning EVH solution Eq.  (\ref{USEVHMetric2}) represents an EVH black hole possessing a non-comapct manifold in $(t,r) = const.$ slices, but with a finite area.

For example in five-dimensional case, the obtained EVHUS metric Eq.  (\ref{USEVHMetric2}) has just one polar coordinate $\theta$, which $\mu_i$s are parametrizing as $\mu_1=\cos\theta$ and $\mu_2=\sin\theta$. Then, in this case the induced metric on the $(t,r)= const.$ slices reads
\begin{eqnarray}\label{MP5-ds2h}\nonumber
ds^2&=&\frac{l^2\,\sin^4\theta}{r^2+l^2\cos^2\theta} d \varphi_1^2  +
r^2 \cos^2\theta d\phi_2^2 + \frac{r^2+l^2\cos^2\theta}{\sin^2 \theta} d\theta^2.
\end{eqnarray}
To explore there is no problem near the pole $\theta=0$ $( 0 \leq \theta \leq \pi/2)$, we use the change of coordinates $k=l(1-\cos \theta)$. Then for small $k$, the induced metric reads
\begin{equation}\label{eq12}
ds^2_h=(r^2+l^2)\left[\frac{dk^2}{4k^2}+\frac{4k^2}{l^2} d\varphi_1^2\right]+ r^2 d \phi_2^2.
\end{equation}
which is clearly describes a metric with constant negative curvature 
\begin{equation}
\mathcal{}{R}=-\frac{8}{l^2+r^2},
\end{equation}
that shows there is no problem near the poles $\theta=0$. That indicate all $(t,r)=const.$ slices has a non-compact topology. This result can be generalized to near-EVH ultraspinning geometry as well. Also for both cases the area of the $(t,r\neq r_+)$ is a finite value. 

It is important to note, to ensure our obtained geometry is free of problem at the poles $\mu_k=0$, one should show that any outgoing null geodesics can not to reach to the $\theta=0, \pi$ axis in any finite affine parameter. It was investigated by detail for ultraspinning Kerr-AdS black hole \cite{1504.07529}.


%

\end{document}